# Gender and Careers in Platform-Mediated Work: A Longitudinal Study of Online Freelancers

PYEONGHWA KIM, Syracuse University, USA
STEVE SAWYER, Syracuse University, USA
MICHAEL DUNN, Skidmore College, USA

We advance gender-inclusive research within the CSCW field by investigating the long-term gendered experiences of online freelancers on digital labor platforms. The prevalence of gender-based inequalities has attracted significant attention within the CSCW community. Yet, insights remain limited on how these inequalities shape workers' long-term experiences on digital labor platforms. Through a five-year longitudinal study of 105 freelancers on Upwork, we reveal persistent gender disparities that influence workers' long-term work and career trajectories, raising concerns about the sustainability of platform-mediated work. We advance the ongoing dialogue on gender inclusivity in the community by introducing the concepts of career disempowerment and platform-mediated motherhood penalty and by offering research and design implications for CSCW to foster more sustainable, equitable platform work environments for all genders.

## 1 Introduction

We contribute gender-inclusive research to the platform work scholarship within the Computer-Supported Cooperative Work (CSCW) community by advancing conceptual and empirical insights into long-term gendered work experiences and consequences on the digital labor platforms. We do this in response to the gender research scholars' call for (1) a longitudinal research looking at gender disparity evolved over time to understand the long-term sustainability of platform-mediated work and (2) its implications for platform work research and design practices to support gender-inclusive sustainability for workers navigating their work and careers on the platforms [38, 54, 70, 76].

### 1.1 Three Discourses on Gendered Platform-Mediated Work Research

This work is motivated by and positioned within three intersecting realities: (1) the dominant presence of female workers on digital labor platforms, particularly in online freelancing, which informs our empirical focus on this phenomenon, (2) theoretical and empirical insights from existing CSCW research on gendered platform work experiences that shape our conceptual and methodological approach; and (3) the movement among CSCW and HCI researchers toward gender-inclusive research that guides our research design and implications for the CSCW and cognate community.

Approximately 47% of the worldwide workforce earns their livelihood through freelancing [39], and 70% of freelancers find their work through digital labor platforms [73]. Women make up 42% of the online freelancing workforce, a larger proportion than the 39.7% in the general labor market [21].

Much of the extant platform work scholarship in the CSCW community tends to focus on microtask or location-based platform labor (e.g., Amazon Mechanical Turk and Uber). This noted, scholarly attention is now turning to online freelancing [2, 44, 45, 54]. This turn arises in part because online freelancing is a “new genre of work" characterized by the need for advanced skills, longer-term project-based work, and a greater degree of spatiotemporal flexibility [11, p.3]. Due to these characteristics and its increasing prevalence in the labor market, scholars view online freelancing as a window into one future of work [9].

Acknowledging this likely future of work, CSCW researchers have voiced concerns about how long-documented inequalities in traditional workplaces are reproduced and may be exacerbated in platform-mediated work [4, 30, 31, 47, 64, 70, 89]. For example, Kim and colleagues' literature review on online freelancing highlights that gender-based inequality occupies the largest space among papers published in CSCW and related fields between 2017 and 2022 [54]. Munoz and colleagues (2024) conclude that digital labor platforms perpetuate traditional gender stereotypes and expectations [70].

Building on research drawn from CSCW and related fields, we frame the theoretical and empirical basis for the work reported here. First, we conceptualize digital labor platforms as "sociotechnical artifacts" in which user experiences are situated within the gendered realities of the settings they are operated on [4, 30, 70, 76]. This perspective acknowledges that the way individuals interact with the platforms emerges at the intersection of the platforms' technological arrangements (e.g., affordances, information architectures, and user experience and interface design) and the prevailing gender inequalities embedded in the societal environments where the platforms are designed and used. Thus, our analytical focus is on exploring how this intersection emerges and evolves, influencing workers’ experiences on digital labor platforms.

We also note that much of the current literature relies on either single-case or cross-sectional studies [54, 83, 91]. Recognizing this issue, both CSCW and gender and technology scholars call for a deeper understanding of the long-term experiences of platform workers to establish a comprehensive understanding of the processual unfolding of their experiences [26, 38, 54, 55, 76, 91].

More broadly, Wajcman (2000) highlights that the analysis of gender and technology requires an integrated approach considering both men and women [91]. Thus, our study focuses on how known inequalities unfold, evolve, and change over time, thereby influencing the long-term livelihoods of all genders on digital labor platforms.

In addressing such inquiry, we align our research with the broader movement toward gender-inclusive research in CSCW. Gender-inclusive research refers to scholarly practices that examine how gender intersects with the diverse behaviors, perceptions, and needs of users interacting with technology, aiming to contribute to the design of technologies that accommodate such diversity [83]. Building on Stumpf and colleagues (2020), we argue that integrating gender-inclusive research into the study of digital labor platforms contributes to the future of platform-mediated work [83]. This integration helps address gender barriers that arise from the interplay of social realities and technological arrangements, thereby fostering equitable opportunities for workers of all genders to succeed in the emerging digital labor market.

### 1.2 Research Questions and Contributions

Motivated by the call for attention from CSCW scholars, informed by the conceptual and methodological insights of existing studies, and aimed at contributing to the broader scholarly discourse of gender-inclusive research, we formulate the following research question (RQ): *How do gendered work experiences shape the long-term trajectories of workers on digital labor platforms?*

To address this question, we employ a longitudinal, mixed-method study of 105 self-identified female and male freelancers on Upwork, one of the largest digital labor platforms. Our findings offer three contributions to CSCW research and practice as summarized here and detailed below.

First, we provide empirical evidence of gendered platform work and career experiences and their evolution overtime. Second, we introduce two concepts that expand CSCW's scholarship on the intersections of gender, technology, and careers in digital labor: (1) career disempowerment and (2) platform-mediated motherhood penalty. Third, we provide both design and research implications for gender-inclusive work environments.

## 2 RELATED WORK

To examine our research question, we draw on two interrelated bodies of scholarship. First, we engage with CSCW and HCI research that investigates the gendered dynamics of platform-mediated work, with particular attention to how algorithmic management systems and platform governance structures shape workers' everyday experiences, often to the detriment of marginalized groups. Second, we build on career studies to understand how these patterned experiences accumulate over time, influencing freelancers' ability to pursue sustainable and meaningful career trajectories. By bridging insights from CSCW and career literature, we develop a conceptual perspective for understanding how daily gendered challenges in digital labor can reverberate over time, ultimately shaping workers' long-term career paths.

### 2.1 Digital Labor Platforms and Gender

Digital labor platforms are privately operated online services that connect demand (employers/clients) with supply (online freelancers/workers) in the labor market. Scholars in both CSCW and cognate communities such as HCI have studied how algorithm- and platform-based control mechanisms reconfigure traditional worker-employee relationships through the concepts of algorithmic management [60] and platform management [50]. The largest digital labor platforms (such as Upwork.com and Fiverr.com) mediate workers' end-to-end labor process including job search, task execution, client communication, and performance evaluation via its own managerial systems.

A growing body of CSCW and HCI literature has illuminated how these platforms often reproduce and even magnify existing social inequalities through technology-driven managerial mechanisms. These include opaque algorithms [60, 68, 77], asymmetrical information architecture ([58], inequitable terms of services [82, 84], and discriminatory user interface design choices [70] among others.

Recently, CSCW scholars have increasingly focused on the gendered implications of digital labor platforms' managerial control, demonstrating that digital labor platforms may not neutral technological intermediaries, but structuring forces that interact with and reinforce gender norms and biases [54, 90, 8, 15, 24, 28, 31, 54, 69, 82, 92].

To better understand this work, we conducted a targeted literature review using the Association for Computing Machinery Digital Library (ACM DL), focusing on publications from 2014 to 2023. The search was guided by keywords including women, gender, platform work, gig work, and platform economy, with selective reviews on studies published in CSCW and related venues. While a systematic review is beyond the scope of this paper, we synthesize findings from our review into two key themes that capture the dominant ways in which gendered experiences in platform-mediated work have been documented: (1) gendered labor shaped by platform design and algorithmic management, and (2) gendered work participation and outcomes influenced by social factors. These dimensions are tightly interwoven, and thus gendered experience are often emerged as sociotechnical issues [13, 22].

The first strand of research highlights the gender-agnostic nature of platform management manifested through two main mechanisms: algorithms and design. Platform algorithms are computational mechanisms that automate the management, organization, and evaluation of work processes on digital labor platforms [60, 67]. These algorithmic processes typically operate in the backend, shaping workflows and decisions invisibly [29]. In contrast, platform design refers to the user-facing interface that renders algorithmic decisions visible and actionable, defining what interactions are possible and how users navigate tasks, opportunities, and evaluations. In other words, while algorithms govern the underlying logic of the work process, design dictates how this logic is materialized and navigated through user-facing features and affordances. The gender-agnostic nature of platform algorithmics and designs often induce gendered experiences. Documented empirical evidence includes: (1) gendered visibility gaps, where women are ranked lower in search results compared to men [38], (2) expectations of constant availability that conflict with caregiving responsibilities, disproportionately disadvantaging women [4, 49], (3) unsafe work conditions due to the lack of harassment and cyberviolence regulations and reporting features [64, 82].

Second, even when platform features are ostensibly gender-neutral, user behaviors and systemic biases contribute to gendered disparities in platform-mediated work. These include: (1) persistent gender pay gaps [31, 42, 70, 74], with these gaps widening further during times of crisis such as the COVID-19 pandemic [15], and disproportionately affecting women of childbearing age [26]. (2) occupational segregation, where women are less likely to be hired for traditionally male-typed jobs [32, 61] and has a dominant presence in stereotypically female-oriented tasks [15]. (3) gendered participation motivations, where women are more likely to engage in platform work due to the appeal of flexible scheduling [16], a factor that is especially salient for mothers [37]. And (4) gendered evaluation and client interaction, including fewer reviews for women [38], lower ratings for women [47], stereotyped appraisals focusing on appearance rather than expertise [20], and exposure to cyberstalking, harassment, and pejorative communications [70, 82].

The literature reveals gendered reality that demands women freelancers constantly adapting: they adjust pricing strategies [30], carefully curate their online presence [70, 76], and seek out supportive networks [64], all in an effort to succeed on platforms not originally built with their realities in mind. Although these adaptive practices showcase workers' resilience, they also raise a deeper inquiry: what are the enduring consequences of navigating such gendered conditions? Beyond the immediacy of day-to-day survival, it is vital to ask whether these cumulative experiences pave the way toward sustainable career paths or instead deepen a trajectory of persistent, and potentially worsening precarity. We now turn to a body of literature that adopts a longitudinal, career-oriented lens to explore how gendered dynamics on platforms evolve over time and shape freelancers' long-term work trajectories and career outcomes.

### 2.2 Digital Labor Platform and Careers

While digital labor platforms are often viewed as one future of work, the question of whether this future is equitable for all genders remains a topic of ongoing debate among CSCW and HCI scholars [9, 30, 38, 47, 64, 70]. To evaluate the viability of platform work for all genders, we must first examine how workers' experiences on these platforms unfold over time, and to understand the implications of these changes for their longer-term career trajectories.

Viewing platform-mediated work through the lens of career theory provides both analytical and conceptual perspective to reveal long-term experiences of workers pursuing their careers on digital labor platforms [52]. For us, a career is "the evolving sequence of a person's work experiences over time" [5, p.8]. A career focus shifts the lens from the more common focus on workplace studies in CSCW research to guide its design practices [63]. A focus on career reflects Barley's argument that "too often CSCW literature does not take advantage of or build on theories developed in the organizational literature, even when directly applicable" [7].

Career theorists have underscored the value of the career concept for its interpretive flexibility, enabling an analysis of a wide range of evolving career phenomena beyond traditional organizational boundaries [5, 17, 66, 75, 85]. This is useful for our research context as the career concept offers a macro-level perspective to comprehend how a series of jobs can translate into work and career trajectories, elucidating how this career shaping occurs in conjunction with technological and gendered contexts. Thus, the career lens provides a dynamic perspective on how work experiences evolve over time, capturing a progression of challenges and successes rather than a static snapshot. This approach allows us to understand gendered work experiences as fluid and continuously shaped by accumulated opportunities and obstacles. Additionally, it highlights how these experiences are interconnected with an individual's social realities, positionality, and the platforms' technological arrangements.

One contribution of a longitudinal approach is identifying the sustainability of platform work as a career. We define career sustainability as a sequence of work experiences that exhibit a pattern of continuity over time, thereby providing positive meaning to the individual [36, 88]. Scholars view career sustainability as a holistic concept that is manifested through the interplay amongst the multiple career sustainability indicators. The literature on career sustainability highlights six key indicators: (1) health (e.g., physical and mental wellbeing) [25, 35, 40, 43, 78, 86] (2) work and life balance (e.g., regular opportunities for rejuvenation and engagement in personal life) [23, 72, 87], (3) happiness (e.g., psychological satisfaction) [25, 35, 43], (4) flexibility (e.g., career evolving to fit an individual's needs and priorities) [14, 57], (5) economic security [56], and (6) employability (e.g., continuous learning and skill development opportunities that improve future career prospects) [25, 27, 41, 59].

Contemporary research is beginning to bridge the gap between CSCW's rich, context-focused, insights on platform work and career theorists' long-range view of work trajectories. Recent CSCW studies now explicitly adopt a longitudinal, worker-centered methodology, tracking the evolution of freelancers' circumstances and strategies [9, 52]. For example, Blaising and colleagues (2020) conducted a longitudinal study with online freelancers, uncovering the hidden "overhead" that workers bear to sustain an online career over time. They found that long-term engagement in freelancing entails a unique set of financial, emotional, relational, and reputational burdens. However, the burdens identified may weigh more heavily on those who also face gender-based hurdles. For instance, if female freelancers must "ask for less" initially to get in the door [31], they start their platform careers from a lower rung, potentially translating into slower earnings growth even as they gain experience [26, 30].

These findings reinforce that time is a critical factor: short-term snapshots may miss slow-moving processes such as the accumulation of advantages (for some) and disadvantages (for others) on digital platforms. Therefore, despite growing scholarly attention, current research offers limited understanding of how these gendered dynamics unfold over time and influence freelancers' long-term career outcomes. Indeed, Hannák and collogues call for longitudinal investigations into gendered trajectories in digital labor [38]. Moreover, translating advantages and disadvantages into the CSCW domain can suggest opportunities for designing socio-technical systems to support platform workers not just in finding the next task, but in building a sustainable future of work.

## 3 METHODS

We present our research approach, beginning with our longitudinal data collection process and the multiple data collection techniques. Following this, we detail our analysis methods used to derive the insights presented in the findings section.

### 3.1 Participant Sampling

We chose to focus on a single platform as an intentional design decision to ensure analytical consistency. Our study centers on Upwork, which the majority of participants identified as their primary and most successful source of work. This platform-specific approach enables us to examine how gender shapes individual experiences within a stable sociotechnical context, minimizing confounding variation that might arise from cross-platform differences.

We used purposive sampling to form our panel, guided by the existing platform work literature [80, 95]. Our selection criteria included workers who were active on the platform, had earned at least $1,000, and were engaged with U.S.-based employers to ensure similar labor regulations and policies. Since Upwork does not disclose demographic data, we invited a large pool of potential participants to secure a diverse sample. Following approved protocols, we contacted each worker, inviting them to participate in a one-hour job. All participants were fully compensated and received a five-star rating with positive feedback.

Drawing from the work of gender and technology scholars [12, 62, 83], we recognize that physiological traits, legal gender, gender expression, and self-identified gender identity may not align. Thus, we included an open-ended text box ("I identify as:") alongside standard gender categories. Of 108 participants, 105 indicated their gender (see Table 1).

Table 1. Participant Demographics

| Category | | Total 108 | |
|---|---|---|---|
| | | Number | Percentage |
| Gender | Female | 63 | 58% |
| | Male | 42 | 39% |
| | Unknown | 3 | 3% |
| Race | Asian | 9 | 8% |
| | Black/African American | 22 | 20% |
| | White | 56 | 52% |
| | Multi-racial | 13 | 12% |
| | Other Race | 6 | 6% |
| | Unknown | 2 | 2% |
| Education | Post-graduate degree | 38 | 35% |
| | Bachelor's degree | 46 | 43% |
| | Associates | 7 | 6% |
| | No college degree | 15 | 14% |
| | Unknown | 7 | 6% |
| Marital status | Divorced | 5 | 5% |
| | Married | 64 | 59% |
| | Never married | 35 | 32% |
| | Separated | 2 | 2% |
| | Widowed | 1 | 1% |
| | Unknown | 1 | 1% |
| Age | 18-29 | 23 | 21% |
| | 30-49 | 70 | 65% |
| | 50-64 | 13 | 12% |
| | 65 and older | 1 | 1% |
| | Unknown | 1 | 1% |
| Occupation | Administrative (e.g., virtual assistants, bookkeepers) | 44 | 41% |
| | Technology (e.g., software developers, data scientists) | 41 | 38% |
| | Creative (e.g., graphic designers, creative writers) | 23 | 21% |

### 3.2 Data Collection

We designed a multi-method approach to data collection. This makes sense in this situation as the different approaches to data collection that we use are complementary. This resulted in developing three datasets during each round of data collection: interviews, surveys and profile archival data. Data for this analysis were provided by online freelancers who have participated in our panel study that has run once each year from 2019 to 2024, with an average interval of 12 months [79].

The five-year timeframe reflects the duration of data collected in our ongoing longitudinal panel study, which began in 2019 following pilot research on freelancing and platform work. While the classification of five years as "long-term" is context-dependent, this period has enabled us to observe temporal shifts in participants' experiences that shorter-term studies may not afford. The reality of a longitudinal study is reflected in the varied levels of participant engagement over time. While some individuals contributed data in only one round, many remained involved across multiple rounds, supporting a robust longitudinal analysis. A total of 105 unique participants took part in the study, with 291 interviews and 327 surveys across five rounds. Among interviewees, 21 participated in one round, 25 participated in two rounds, 21 in three, 13 in four, and 21 in all five rounds. Among survey respondents, 27 participated in one round, 20 participated in two rounds, 21 in three, 18 in four, and 25 in all five rounds.

*3.2.1 Interviews.* For every round of data collection, each freelancer participated in a 45-minute interview. Interviews focused on three themes. The first focus examines how workers define and interpret work and career sustainability within the context of platform work. The second area of focus explores how gender intersects with technological, social, personal and other related factors to shape individuals' work experiences and long-term careers. To explore this, we incorporated probing questions about what supports or hinders freelancers in sustaining their work, both in the short and long term. Third, we examined changes in both professional and personal contexts, including aspects such as family status, employment situation, to understand how these relate to workers' evolving perceptions and experiences of sustainability on the platform.

*3.2.2 Surveys.* Designed as a companion to the interview, each freelancer completed a 15-minute and online questionnaire in every round of data collection. The survey focused on six themes: motivation for participating in platform work, work and employment strategies, work behaviors, sociodemographic status, platform work outcomes and career plans.

*3.2.3 Profiles Data.* We reviewed data from each participant's public platform profile. Each profile includes both quantitative data (e.g., job success scores, accumulated work hours measured by the platform, client ratings) and qualitative data (e.g., professional skills and educational qualifications, project histories and client reviews).

### 3.3 Data Analysis

We pursued two lines of data analysis. We provide a descriptive analysis of our survey data consisting of both numerical and categorical data. This analysis aims to identify any notable differences or trends between these groups, providing insights into gender-based variations within among our research participants.

For the survey analysis, we included data from participants who provided valid responses to each survey item. The dataset was carefully examined to detect and address missing values and outliers. Missing data, such as responses marked "prefer not to answer" and other non-responses, were excluded from the analysis. Outliers, defined as observations that substantially deviated from the overall data distribution, were also removed to ensure analytical accuracy. Descriptive statistics were calculated for numerical variables, including the mean, median, standard deviation, and range. For categorical variables, frequencies and percentages were reported. Variations in the number of respondents across survey items and rounds reflect the typical nature of panel studies, in which participant engagement may vary over time, and not all individuals respond to every item in each survey wave.

For the interview analysis, we employed a grounded theory method and thematic analysis [10, 33, 81]. Considering the diverse characteristics of the dataset and the longitudinal nature of data collection, this study follows practical guidance for thematic exploration [94]. The thematic analysis began with the first author open-coding 10% of the interview data and sharing initial notes in weekly team meetings to discuss emerging keywords and descriptions. Preliminary codes were then applied to an additional 15% of the data, identifying recurring themes and new examples that helped draft an initial coding manual. In continued weekly meetings, the team organized these codes into broader categories, laying the foundation for axial coding and resulting in a codebook with 22 codes, including career changes, obstacles, gender roles, and platform features.

Using this codebook, the first author coded the full dataset. The coding was done with NVivo 14. The profile data were coded by hand. This process produced 10 themes, (e.g., spatiotemporal flexibility, career shifts, and platform impacts on career sustainability). All authors and five regular team members refined these themes through weekly discussions, resolving disagreements until reaching consensus on the final themes reported in this paper.

## 4 FINDINGS

Data analysis makes clear there are gender disparities in the experiences of workers on the platform, with women facing more challenges. In the rest of this section, we present five themes derived from survey data, and three themes from interview data, that help make clear these gendered disparities.

### 4.1 Survey Findings

Survey results show both female and male participants have increasingly moved away from viewing platform work as a sustainable career path. This decline is more pronounced among females, with fewer considering it a viable long-term option as compared to males. Data also reveal that both genders have shifted toward seeing platform work as primarily for part-time arrangements.

Female participants have experienced a sharper reduction in weekly work hours and more frequent but less successful bids for platform jobs compared to males. Evidence shows that over the time of this study, the earnings from platform work have declined for both genders. And, while both report decreasing monthly earnings, males have experienced a steeper decline over time. Finally, data show that females primarily enter platform work due to domestic responsibilities, while males are mainly driven by financial obligations, with more males serving as primary breadwinners compared to females.

*4.1.1 Platform Work as a Career Path.* Both female and male participants exhibit a significant decline in viewing platform work as a long-term career. Among female participants, 74% considered platform work part of their long-term career plan in Round 2 (2019). The figure dropped sharply to 38% by Round 5 (2024). Notably, the percentage of females not actively pursuing new jobs increased from 0% in Round 2 to 42% in Round 5. Similarly, 65% of male participants initially viewed freelancing as a long-term plan in Round 2. However, by Round 5, this figure dropped to 43%. Like the females, the proportion of males not actively seeking new jobs rose from 0% in Round 2 to 22% in Round 5.

Table 2. Platform work as a career path

(%)[2]

| How do you currently view Upwork? | Gender | R2 | R3 | R4 | R5 |
|---|---|---|---|---|---|
| Long-term plan | Female | 74 (N=39) | 67 (N=36) | 67 (N=30) | 38 (N=24) |
| | Male | 65 (N=26) | 63 (N=32) | 67 (N=27) | 43 (N=23) |
| Not sure about how long I will stay on Upwork | Female | 13 | 19 | 23 | 4 |
| | Male | 23 | 19 | 19 | 22 |
| For the next few months to a year | Female | 8 | 8 | 3 | 4 |
| | Male | 8 | 9 | 11 | 13 |
| Not actively pursuing new jobs currently | Female | 0 | 0 | 0 | 42 |
| | Male | 0 | 0 | 0 | 22 |
| Have not really thought about this | Female | 5 | 6 | 7 | 13 |
| | Male | 4 | 9 | 4 | 0 |

*4.1.2 Platform Work Engagement.* We tracked indicators of workers' engagement in platform-mediated work to assess how their involvement evolves over time, focusing on: working arrangements, weekly working hours, and the number of project bids submitted and accepted.

Table 3. Employment Status

(%)

| How do you classify your current employment status on Upwork? | Gender | R2 | R3 | R4 | R5 |
|---|---|---|---|---|---|
| Part-time | Female | 71 (N=31) | 70 (N=37) | 82 (N=50) | 71 (N=28) |

[2] The displayed percentages represent the proportion of the total female or male respondents in each round who selected each survey item. For example, in R2 74% of the total female respondents indicated that they view Upwork as part of their long-term plan.

| | | | | | |
|---|---|---|---|---|---|
| | Male | 30 (N=19) | 79 (N=34) | 81 (N=32) | 76 (N=25) |
| Full-time | Female | 26 | 19 | 14 | 11 |
| | Male | 35 | 12 | 19 | 24 |
| Other (e.g., Upwork is helping me close career gaps during my job search.) | Female | 3 | 11 | 4 | 18 |
| | Male | 17 | 9 | 0 | 0 |

Table 3 makes clear that both genders are increasingly moving toward part-time work.

Table 4. Weekly Working Hours

(Hours)

| How many hours per week are you working on Upwork? | Gender | R1 | R2 | R3 | R4 | R5 |
|---|---|---|---|---|---|---|
| | Female | 9 (N=39) | 6 (N=32) | 5 (N=38) | 2 (N=48) | 1 (N=26) |
| | Male | 5 (N=26) | 5 (N=22) | 5 (N=35) | 5 (N=35) | 5.5 (N=23) |

Table 4 provides a summary of female participant's decline in their weekly platform work hours compared to males. In the initial round of 2019, women reported a median of nine hours per week dedicated to platform work. By the fifth and most recent round in 2024, this had sharply decreased to a median of just one hour per week, marking an eight-hour reduction. In contrast, male freelancers maintained a relatively consistent weekly work schedule, around five hours across all rounds.

Table 5. Weekly Job Bid Submission and Accepted bids

(Number of bid)

| How many bids on Upwork do you submit in a typical week? How many of these proposals for Upwork jobs get accepted? | Gender | R1 | R2 | R3 | R4 | R5 |
|---|---|---|---|---|---|---|
| | Female | 2 (N=31) | 2 (N=27) | 3 (N=25) | 3 (N=16) | 4 (N=26) |
| | Accepted | N/A | 0.6 | 0.8 | 0.26 | 0.5 |
| | Male | 5 (N=28) | 3 (N=24) | 5 (N=27) | 4 (N=16) | 2 (N=25) |
| | Accepted | N/A | 0.8 | 1.5 | 0.8 | 0.5 |

Table 5 contains a summary of weekly bid submissions and acceptance rates for female and male participants across five rounds. The data reveal a steady increase in weekly bid submissions by female participants, while male participants display more variability, leading to an overall decline. Despite higher submission activity, female participants saw a decrease in accepted bids, whereas male participants also experienced a decline but maintained consistently higher acceptance rates than females.

*4.1.3 Monthly Earnings:* Table 6 contains a summary of freelancer's monthly earnings. This helps make clear the decline in median monthly income on the platform is more pronounced for men than for women. From Round 1 to Round 5, women's median monthly income decreased by $800, falling from $900 in Round 1 to $100 in Round 5. In contrast, men's median monthly income declined by approximately $1,025, from $1,300 to $275 during the same period.

Table 6. Monthly Earning

($)

| In a typical month, approximately how much do you earn via Upwork? | Gender | R1 | R2 | R3 | R4 | R5 |
|---|---|---|---|---|---|---|
| | Female | 900 (N=47) | 350 (N=30) | 350 (N=28) | 100 (N=32) | 100 (N=27) |
| | Male | 1,300 (N=28) | 700 (N=22) | 500 (N=26) | 400 (N=24) | 275 (N=24) |

*4.1.4 Drivers for Pursuing Platform Work and Breadwinner Status.* Data in Table 7 highlights gender differences in motivations for entering platform work. Financial reasons were a primary driver for 52% of male participants, compared to 32% of females. Family responsibilities were cited by 33% of female participants but only 5% of males. Career-related motivations were important for both genders, slightly more so for males (36%) than females (29%). Flexibility was rarely mentioned, with only 7% of males and 6% of females citing it as a main reason for starting platform work.

Table 7. Main Motivations for Entering Platform Work

(%)

| What made you start working on Upwork? | Gender | R1 |
|---|---|---|
| Financial (e.g., to secure primary or secondary sources of income) | Female | 32 (N=39) |
| | Male | 52 (N=28) |
| Family (e.g., to take on caregiving responsibilities for children or family members) | Female | 33 |
| | Male | 5 |
| Career (e.g., to transition into new career paths or develop new skills) | Female | 29 |
| | Male | 36 |
| Flexibility (e.g., spatial or temporal freedom or lifestyle preferences) | Female | 6 |
| | Male | 7 |

As shown in Table 8, across all rounds of data collection, female participants reported being the primary breadwinner at a lower rate compared to their male counterparts. Conversely, over 50% of our total self-identified male study participants consistently identified themselves as the primary breadwinner across all rounds. The data indicates that men are more likely to be the primary breadwinner.

Table 8. Breadwinner Status

(%)

| Are you considered your household's "primary breadwinner"? | Gender | R1 | R2 | R3 | R4 | R5 |
|---|---|---|---|---|---|---|
| | Female | 36 (N=39) | 25 (N=31) | 69 (N=35) | 49 (N=29) | 27 (N=25) |
| | Male | 58 (N=28) | 68 (N=22) | 81 (N=33) | 67(N=23) | 54 (N=25) |

## 4.2 Interview Findings

This section presents three findings from analyzing the interview data regarding the differential experiences of female and male participants: (1) gendered experiences as enablers of career sustainability, (2) gendered experiences as constraints on career sustainability, and (3) the evolving nature of career sustainability shaped by gendered dynamics.

*4.2.1 The Enabler of Career Sustainability: Flexibility for All, with Differences between Women and Men.* Flexibility is a cornerstone of sustainable careers on digital labor platforms for all genders. Both self-identified female and male participants highlight flexibility as a critical factor that elevates platform work from a mere temporary stopgap to a legitimate, long-term career option capable of providing stability for several years, or even an entire career.

*Spatial, Temporal, and Task Flexibility as Enablers of Female Freelancers' Career Sustainability on Digital Labor Platforms.* The platform's spatial flexibility supports remote work needs that many female participants view as essential for sustaining long-term careers on digital labor platforms. Upwork operates on a remote-first model, offering project opportunities that surpass geographic constraints. This setup allows female participants to stay active in the workforce even when relocating to areas with limited quality jobs, insufficient local client networks, or where their professional credentials hold less value.

While the flexibility offered by digital labor platforms can benefit freelancers of all genders in such situations, our findings reveal a distinctly gendered experience. When platform-enabled flexibility intersects with the career trajectories of female participants, it becomes crucial for their career sustainability, as many female participants face the challenge of involuntary relocation. Involuntary relocation is a recurring theme in the narratives of female participants, as they often move to follow their partners' careers.

First, female participants frequently emphasized the role of platforms in providing the technological infrastructure that enables them to sustain their careers after relocations. Notably, only self-identified women recounted these experiences. Participants P15, P17, P19, and P33, all of whom hold tertiary degrees, had previously worked in traditional labor markets. However, after relocating to support their partners, digital labor platforms became their primary, and often sole, means of maintaining a professional career.

> "We moved to Illinois for his career. I have a master's in education, but I'm not certified to teach in Illinois. [...] I've been able to pursue educational writing jobs and educational clients, which is really lovely [...]. I probably will stick with it, only because I live very rurally, so well-paying jobs are just not available where I am, jobs that utilize my skills are just not really available where I am". P19 (Female freelancer)

> "We moved to the Seattle area because of my husband's job. I did not do any full-time jobs here. I don't have any clientele here. Upwork is like the perfect thing for that scenario". P33 (Female freelancer)

Second, the platform's temporal flexibility plays a crucial role in shaping female participants' views on career sustainability, especially for those balancing dual responsibilities: work and domestic duties. Upwork supports this flexibility by offering detailed job postings that specify key attributes, like project type (one-off or ongoing), which indicate the expected time commitment. These details help freelancers select projects that fit their temporal availability and personal constraints.

For female participants, this temporal flexibility is not just a convenience but a necessity for their career to be sustainable. As P67 explains, although she and her husband both work remotely on Upwork as part-time freelancers, their domestic responsibilities differ. Consequently, the female partner in a dual-career couple places considerable importance on the temporal flexibility afforded by digital labor platforms, considering it essential for the sustainability of her career.

> “Managing household arrangements while working is very hard. My husband and I are both full-time volunteers and part-time freelancers. Then, I take on the bulk of our housework and caregiving. So, that is part of why freelancing appealed to me because I could set my own hours”. P67 (Female freelancer)

Similarly, P25, a mother of young children, emphasizes that the ability to adjust her work hours around caregiving tasks is essential for balancing professional and domestic responsibilities. P18, caring for elderly parents and a child with special needs, also relies on this flexibility, as traditional work hours would make employment incompatible with her caregiving duties. For female participants, temporal flexibility is crucial, allowing them to structure their professional lives around domestic roles. This ability to balance both responsibilities is central to their view of a sustainable career.

> “I started on Upwork when I was on maternity leave with my first kid. I had three months’ maternity leave, and I was just looking to make a little money on the side. Two years later we had my second daughter, and just the childcare costs just made sense for me to stay at home and watch them. [Upwork] is a permanent situation for me. I do like the flexibility of staying at home”. P16 (Female freelancer)

> “[Upwork] is long-term. I turned to freelance and gig work after my husband’s death. I was also responsible for the care of my father-in-law, my husband’s father. I also have an adult child who has Asperger’s syndrome and was dealing with a lot of anxiety. So I was not in a position where I could work outside of the home, because I was also a caregiver and needed my time to be free for my family. I envision myself as being a gig worker and a freelancer for the rest of my working life”. P18 (Female freelancer)

Third, task flexibility emerged as a key factor supporting sustainable careers, especially for female participants with caregiving responsibilities. This flexibility allows women to remain active in the labor market while managing domestic and caregiving duties. Upwork offers knowledge work opportunities with details on task complexity (from simple to complex projects) and required skill levels (from entry-level to expert). This task flexibility enables female participants to select jobs that align with the skill levels they wish to use while balancing domestic responsibilities.

The female participants view this arrangement as essential to the long-term viability of their careers (P76, P80, P86, P10), with some citing it as a decisive factor in choosing online freelancing as a permanent path (P81, P85, P86). They describe career sustainability as the ability to balance roles as both mothers and professionals. As P86 emphasized, they appreciate how digital labor platforms allow them to navigate the "dual parts of their careers," with one facet dedicated to professional development and the other centered on meeting family responsibilities.

> “I was a stay-at-home mom then, but at the same time, you just want to do other things than just be a mom. I wanted to do something to improve myself.” P76 (Female freelancer)

> "I need to be available as a caregiver, but still be able to have a career. I've managed to accomplish that. I went into freelance because I had to be available as a caregiver and I still needed an income, so as long as the two can work together, I'm comfortable." P10 (Female freelancer)

> "It does serve me because I have that flexibility. My hundred percent focus is creating time for the family, my kids [...] and then I want to challenge myself, and that freelancing space does allow me to do that." P86 (Female freelancer)

*Work Flexibility as an Enabler for Men Augmenting Autonomy***.** Male participants also view flexibility as a central factor that enables platform career sustainability. They tend to see flexibility as a much broader concept, encompassing location, time, and job structure, and use this freedom to enhance their autonomy in shaping career trajectories according to their professional goals and personal preferences.

For example, P47 illustrates how flexibility in his freelance career allows him to make deliberate choices about how to allocate his time between work, rest, and family, tailoring his schedule according to his personal preferences. This level of autonomy reflects the broader experience of many male participants, who often have the freedom to determine how they use flexibility to suit their individual needs. In contrast, female participants frequently lack the same degree of agency, as they are more likely to structure their work around domestic responsibilities and caregiving duties, leaving them with less flexibility to prioritize their own preferences or personal well-being. We note that when male participants discuss career sustainability in relation to flexibility, there is a consistent absence of references to domestic responsibilities, even among those with children and partners (P42, P47, P49, P55, P83, P85).

> "I just schedule my time as if I work in the office. I start working at 10 o'clock in the morning until about 5 o'clock in the evening as a routine. [....] I'm living with my wife, so she understands that I have to work on a firm schedule. For the day, if there is no assignment or job coming in, I will help her doing the housework, but usually when I have a job to do, she will do the housework". P55 (Male freelancer)

*4.2.2 The Challenge of Career Sustainability: Constant Availability for All, with Differences between Women and Men.* Digital labor platforms demand constant availability, impacting participants' immediate work experiences and long-term career sustainability. Although this requirement applies to all workers, it has distinct gendered effects. Female participants, often balancing caregiving duties, face challenges in meeting these demands. Our analysis identified three main areas where constant availability pressures disproportionately impact female participants: (1) disruptions during critical life events due to platform notifications and client expectations, (2) limited access to quality jobs due to restricted work hours, and (3) reduced visibility and career prospects affected by algorithms and ratings. In contrast, male participants experience pressures of overwork and constant availability, striving to secure jobs and manage client interactions, which also undermines their career sustainability.

*The Burden of Constant Availability and Its Impacts for Women's Career Sustainability.*

*Disruption during critical life events:* Female participants face significant challenges in managing platform-imposed availability during critical life events such as childbirth or caregiving. Interview data revealed that women often struggle to maintain constant connectivity during these periods, which can lead to lost opportunities and long-term career instability (P2, P6, P15, P26, P66). They reported that constant availability is a major challenge, forcing them to work under a "fear of losing clients" (P26), which leads them to feel they need to be "responsive and available a little bit around the clock" (P2). Female participants are continuously checking emails and notifications from platforms amid their caregiving responsibilities. For example, P6 described how, even immediately after giving birth, she felt compelled to respond to clients from the hospital.

> "It's almost impossible for me to take time off. The funny thing is, while freelancing I had a baby, and so right after the baby was born, I was in the hospital responding to clients on my phone. So there really is no time off." P6 (Female freelancer)

*Limited access to better-paying, longer term, and higher-skilled work:* The constant availability required by digital platforms also exacerbates the barriers women face in accessing better-paying, longer-term, and higher-skilled work, which then eventually compromise their career sustainability on digital labor platforms. Female interviewees reported that their caregiving responsibilities often restricted the time they could dedicate to platform work, thereby limiting their ability to take on more lucrative projects that typically required long-term engagement (P26, P35). P16 articulated this issue, explaining that her availability was constrained by her children's school schedule:

> "I mentioned childcare. I have a five- and a seven-year-old who just started school, so I'm mainly only available when they're in school, which is like 8:30 to 3:30. [...] I wish I could be getting into more technical work, so there's concern that there's not a lot of that available hourly technical work to match what I have in terms of time."P16 (Female freelancer)

This finding highlights how platform-driven demands for constant engagement favor freelancers who are more flexible with their time, typically those without caregiving responsibilities. Women, who often carry the dual burden of paid and unpaid labor, struggle to compete for high-paying jobs that require full-time availability.

The gendered reality for female freelancers often limits them to specific types of job opportunities, typically low-paying, short-term, low-skill tasks that require minimal cognitive investment or time commitment. One participant, a mother of young children, shared that she typically engages in simple, repetitive tasks while looking after them:

> "My work, it is a no-brainer. It's not particularly challenging, taxing, or exhausting. You just collect your junk mail, white out the information, scan it in, and get paid." P20 (Female freelancer)

*Penalized by platform algorithms and rating systems*: A critical concern for women on digital platforms is the impact of visibility and rating systems, which can negatively affect career prospects if constant availability is not maintained. Female participants are aware that inactivity on the platform may cause them to lose premier statuses such as "top-rated freelancer" or "rising star" (P20, P23, P59). Platform algorithms are designed to prioritize freelancers who engage frequently, making it difficult for women to maintain a competitive edge if they need to take breaks. Women who take time off for caregiving or personal reasons often experience a decline in visibility, which can result in fewer job opportunities and lower ratings.

P20, who raised young children while freelancing online, explained her anxiety over how periods of inactivity affected her top-rated freelancer status:

> "I was a top-rated freelancer. I should say that I had that title until I took a hiatus. I don't know if it's just I didn't work hard enough for employers for a set period of time or how I lost that title." P20 (Female freelancer)

P23 also expressed uncertainty about whether taking time off would lead to the loss of her top-rated status, questioning the sustainability of her career on digital labor platforms:

> "I'm a top-rated freelancer. I don't know how taking any time off affects being top rated. I started doing Upwork when my first child was like a year old. When I was pregnant with my second child, I was doing like an hour or two a week. [...] When I envision having three children, I don't know if that's sustainable, but then if you take time off, I don't know if you can keep your top-rated status. I'm like, 'Will it be easy to get back on the platform, or would I lose my status? [...] It would be nice if they had some sort of maternity leave option, just a grace period" P23 (Female freelancer)

*The Burden of Constant Availability and Its Impacts for Men's Career Sustainability.* As reported above, female participants generally experience constant availability as a constraint that hinders their career progression. Faced with the challenges of balancing domestic and professional responsibilities, they often prioritize caregiving or attempt to divide their efforts equally between home duties and freelance work on platforms like Upwork.

In contrast, male participants exhibit a more intentional approach to managing platform-mandated availability. This is especially evident in their active use of platform features like the Availability Badge, which signals their readiness to take on work on the platform. For example, one participant explained how he strategically uses the availability badge to enhance visibility, signaling that he is constantly connected to potential clients and job opportunities.

> "I do that so people can contact me so I immediately reply to them [...] That's why I turn it on." P65 (Male freelancer)

This deliberate approach to meeting the platform's availability requirements is particularly evident among male participants who serve as the primary breadwinners. These individuals feel an increased obligation to remain constantly available in order to secure long-term projects and foster relationships with repeat clients. As P39 noted:

> "I find that I have to work as much as I possibly can. I have to be available day and night. [...] We would have my son in daycare at least some number of hours per week so that my wife would have more time to do whatever she wanted to do, whether that is housework or work or just relax". P39 (Male freelancer)

Similarly, P61, a male participant supporting his family, highlighted the competitive nature of the freelance market and the necessity of preparing for future work by constantly remaining available:

> “In my family, we are me, the main person who is earning more, and the entire family is dependent on me. My family members also support me in other ways. But now, no, I am the main person who is earning more. [...] The market is very competitive. If I don’t have a job, I would start exploring new clients. I have a mobile app. There is a notification coming in”. P61 (Male freelancer)

*4.2.3 Evolving Perceptions of Sustainability on Digital Labor Platform by Gender.* This section presents findings on how participants' perceptions of the long-term suitability of digital labor platforms evolve over time, with a focus on the gendered experiences that shape these shifts. A central finding is the transition from initial optimism to a more constrained view of freelancing's long-term viability across all genders. This shift is shaped by gender-specific challenges, summarized in four key findings. The outlook on career sustainability becomes less optimistic for all genders due to (1) the financial undervaluation of their work, (2) a diminished sense of career identity and limited career fulfillment. Similarly, men’s perspectives shift as they confront (3) income instability and (4) intense competition and excessive workloads and the resulting burnout.

*The Financial Undervaluation among Women:* The data reveals that freelancers often begin with optimism, viewing freelancing as a viable long-term career. However, many reassess its sustainability over time.

P9’s (female) case illustrates this shift among female participants who initially pursued freelancing to balance domestic responsibilities but later encountered perceived unfair compensation (P16, P22, P26, P27, P28, P107). P9 started freelancing after her company eliminated her position during maternity leave. Reflecting on her decision, she said, “I just decided I’m not going to put a bunch of effort and use my talents and skills for another company that basically doesn’t get punished for pregnancy discrimination.” With over seven years of experience, she turned to Upwork, hoping freelancing would lead to stable work.

In 2019, she was optimistic: "[Freelancing] is stable and sustainable. I definitely plan to grow, and when I get to the point where I can’t handle the work that I’m bidding on, then I’ll have to hire help or something". Despite facing unfair compensation early on, she continued, as it provided essential income for her children’s daycare. By our third interview in 2023, however, her optimism had waned:

> "Freelancing has become less sustainable. When I was at [company name], they charged their clients $65 an hour for anything I did, and there’s no way I could get that as a freelancer right now. I feel like the clients on Upwork are expecting to get something really cheap and quick based on the jobs they post and the offers they send out, so the pay would have to increase.”

Her Upwork profile shows an hourly rate of $22 to $30 from 2019 to 2024, less than half of what she earned at her previous company.

*Career Disempowerment among Women.* Longitudinal data show that female freelancers experience limited professional growth, confidence in their career over time. Despite their professional qualifications and previous corporate experience, they found that their freelancing work became fragmented into temporary, lower-skilled tasks. This fragmentation hindered their career development, career progression, and sense of professional identity.

P16 (Female), a mother of two with a postgraduate degree and ten years of corporate research experience, began freelancing during her maternity leave to keep her skills current while managing childcare. She initially appreciated the flexibility freelancing provided: *"[Having flexible working hours] is nice, especially with the kids. If they're sick or they have an appointment, I can usually work around that."*

However, in 2022, she reflected on the limitations of freelancing and the constraints it placed on her ability to engage fully in her professional field. She described the challenges of balancing caregiving, domestic responsibilities, and freelancing, all of which hindered her capacity to take on additional work and fully leverage her skills.

> "I thought I could get these gig work hours in during the school day and then be available for my family when they need me after school. The volume of my freelance work has gone down, but that's partially because I just have other commitments with childcare. I thought I'd have more time once the kids went back to school, but that hasn't been the case. I'm not able to take on more work."

Since our 2022 interview, no new jobs were recorded on her Upwork profile other than her participation in our research.

P33, an interactive designer with a postgraduate degree and eight years of experience in traditional corporate roles, also transitioned to freelancing to be more available as a mother to her young children. During our initial interviews (2019–2020), she worked on larger design projects, such as entire eBooks. However, by 2022, she had shifted to smaller tasks, such as creating individual graphic elements for larger projects, and viewed freelancing as more of a job than a career.

> "I don't feel like I still have a career-career. I'm just doing this on the side whenever I have time. Maybe in five years' time, when my child is a lot older and doesn't need me as much, I will move back to more of a career. [...] I only look for work when I'm available."

In our most recent interview, she reported that she had not undertaken any new jobs on the platform, mirroring the trajectory observed with P33.

Similarly, P23 (female), a former data analyst/project manager with a master's in IT, left her corporate job to freelance after her first child. Over the next four years, she took on projects ranging from $5 tasks to $4,500 assignments. In our 2019 interview, she shared her motivation:

> "When my daughter was a year old, she started taking these regular, long naps in her crib. For a week or so I just kind of sat around and enjoyed it, but then I felt like I needed something more exciting to do while she's napping."

Over the years, she took on a wide range of roles, including data labeler, proofreader, and virtual assistant, most of which required minimal cognitive engagement and time commitment. By 2023, she continued in these roles and noted:

> "[Online freelancing] has always been pretty limited since I've been balancing it with staying home with my children. There have been times when I've tried to get more clients, make more money, or work more hours, but I've mostly just kept it alive with the bare minimum."

She now primarily identifies as a stay-at-home mother rather than a professional in her field: *"When somebody asks what I do for a living, I just say I'm a stay-at-home mom."*

*Income Instability among Men.* The trajectories of male freelancers reveal a shift from an initially positive outlook on freelancing to questioning its sustainability, largely due to financial compensation not meeting their expectations as primary earners.

P40's experience illustrates this shift among several male participants (P42, P45, P47, P52, P53) who entered freelancing out of "financial necessity." After seven years balancing part-time freelancing with a full-time legal job, P40 transitioned to full-time freelancing, seeking flexibility and income growth to support his household.

In our initial interviews (2019-2020), P40 consistently expressed optimism about the sustainability of his work. In the early interviews, P40 expressed optimism, saying, "*I have a lot of work, and projects coming out the wazoo. So it's sustainable."* By 2020, he added, *"It is quite stable right now. I'm hoping it's sustainable."*

However, by our third interview in 2021, P40 reported experiencing fluctuations in his project flow, which adversely affected his financial goals: *"It certainly raises concerns about the longevity of this work. I'm now more willing to take on extra clients and consider diversifying my income."*

In 2023, P40's concerns were validated by a significant decline in the volume of work he received. This prompted him to reassess his income sources:

> "I think that the pace I'm currently maintaining is not sustainable, but it's not intended to be. I'm in the process of making a significant push right now, moving more towards establishing a full-scale agency rather than focusing solely on freelancing."

*Intense Competition, Workload Management, and Burnout among Men.* As their freelance careers progressed, male freelancers increasingly reported difficulties in navigating the pressures of intense competition and managing excessive workloads in order to avoid burnout. These experiences gradually shifted their perspectives on the long-term viability of careers on digital labor platforms.

In our third interview with P47 (Male), he emphasized the unsustainable nature of freelancing as competition intensified, particularly as the digital labor market became increasingly saturated with new freelancers. He noted: *"There are so many freelancers vying for the same positions. The rates are dropping as more people come on board. It's harder to stand out."*

By the third year, P47 reported significant work fatigue: "I'm just slammed with work for too long and I can't go. The longest I've done is like two months without a break, maybe more."

Similarly, P39 (Male) expressed frustration with the overwhelming volume of work. During our third interview, he described how managing multiple clients simultaneously led to burnout: *"I get overwhelmed with too many projects and end up burning out."* His struggle to balance the demands of numerous projects highlights how workload volume became a major source of stress.

By the fourth interview, P39 was actively seeking ways to better manage his workload, although he recognized the challenge posed by the urgency of client demands: *"I'm trying to* *manage my workload better, but it's tough when clients want everything done yesterday. I might need to cut back to avoid burnout."*

## 5 DISCUSSION

Building on the well-documented gender disparities in platform work within the CSCW community, the research reported here addresses the main question: How do gendered work experiences influence the career sustainability of workers on digital labor platforms? The longitudinal approach reveals that gender disparities evolve over time across different aspects of platform work, impacting career sustainability. Based on our findings we discuss: (1) collective challenges in sustaining platform-mediated careers across genders (2) the gendered nature of platform-mediated work experiences with particular emphasis on two novel concepts: career disempowerment and the platform-mediated motherhood penalty, and (3) the implications of these insights for CSCW.

### 5.1 Shared Long-Term Challenges in Platform Careers Across Genders

Findings provide empirical insight into what it actually means to pursue a career on digital labor platforms—a question that has been a consistent theme of exploration yet remains largely unanswered within the CSCW and HCI communities [65]. Rather than functioning as empowering infrastructures for sustainable employment, digital labor platforms are increasingly associated with a shared, long-term trajectory of declining professional legitimacy, economic instability, and precarious career development across genders. We identify three trends that affect freelancers across the gender spectrum. First, shifting career perceptions: freelancers are increasingly abandoning the view of platform work as a viable long-term career, reflecting unmet expectations for professional growth. Second, decreasing engagement: many are reducing their hours or disengaging altogether, signaling a broader decline in confidence in platform work's sustainability. Third, declining income security: earnings have steadily eroded across genders, highlighting growing financial instability and questioning the long-term viability of platform-based careers.

Framed through a career lens, this convergence of experience allows us to assess platform work using established indicators of career sustainability, such as economic security [56], career growth opportunities and meaning [25, 27, 41, 59]. Instead of viewing platform work solely as a site of individualized, algorithmically mediated transactions, we argue for a reconceptualization of platform work as a sociotechnical experience characterized by collective degradation over time. Addressing this requires reorienting our analytic gaze from micro-level task dynamics to macro-level career trajectories.

While CSCW has made strides in examining algorithmic management and platform precarity [50, 96], studies often focus on short-term dynamics or specific episodes of work. Our longitudinal approach reveals how platform systems may accumulate disadvantages over time, affecting freelancers of all genders, though with a disproportionate burden on women. This aligns with Blaising and colleagues (2021), who describe the "overhead" workers must absorb, financial, emotional, relational, and reputational, to sustain platform-based careers [9].

To advance this discourse of long-term challenges, we encourage CSCW and HCI scholars to more directly engage with how platforms structure freelancers' evolving sense of professional identity, economic value, and long-term viability. These dynamics are not merely individual outcomes but are collectively shaped by platform design, market saturation, shifting algorithms, and broader social conditions. Addressing these issues requires moving beyond static snapshots of gig work toward a systemic, temporal, and collective analysis of platform labor.

Future research should therefore ask: How can platform design shift from optimizing short-term task flows to supporting long-term worker development, retention, and well-being for workers of all genders? Addressing this question not only offers design pathways for more equitable and sustainable digital labor systems but also expands CSCW and HCI's theoretical and methodological commitments, shifting the focus from momentary interactions to the cumulative, structural conditions shaping workers' lives over time.

### 5.2 The Gendered Experiences of Platform-Mediated Work

Our research also contributes to the platform work scholarship in CSCW, particularly its growing debate on the role of gender in shaping platform work experiences [8, 15, 24, 28, 31, 54, 69, 82, 92]. While these studies shed light on how platforms amplify traditional labor market inequalities, few examine how these experiences impact the sustainability of platform work as a viable career over time. Building on existing insights, our longitudinal approach reveals how gendered platform work experiences evolve and compound over time, shaping workers' trajectories in ways that are not immediately visible in short-term analyses.

We discuss three insights that clarify the contours of these long-term gendered platform work experiences. First, we demonstrate that motivations for engaging in platform work are gendered: women often enter platform work to accommodate caregiving responsibilities, whereas men are more commonly motivated by financial imperatives, reinforcing conventional gendered divisions of labor. Second, we conceptualize career disempowerment to articulate how platform-enabled flexibility, while facilitating short-term labor participation, simultaneously constrains women's long-term career sustainability. Third, we introduce the concept of the platform-mediated motherhood penalty to explain how algorithmic visibility mechanisms and engagement metrics disproportionately penalize mothers for caregiving-related absences, thereby diminishing their professional standing and access to future opportunities. Collectively, these contributions extend CSCW scholarship by foregrounding how platforms' technological arrangements intersect with gendered realities to shape uneven career trajectories.

*5.2.1 Gendered Motivations for Entering Platform Work: Caregiving vs. Financial Considerations.* Findings make clear that the motivations for engaging in platform work are gendered, suggesting that platform work participation continues to mirror traditional gender divisions of labor: women tend to pursue platform work as a means of accommodating traditionally gendered responsibilities such as caregiving, while men do so to fulfill breadwinner roles [1, 93]. This observation is consistent with existing research on gender-based participation in platform work, as noted in the literature review [16, 37, 51, 71]. Our research advances these findings by demonstrating how the gendered division of labor manifests in platform work participation and its long-term sustainability on these platforms.

In Round 1 of our study in 2019, we asked participants how they transitioned into online freelancing on the platform. This question was followed by inquiries about their paid work experiences prior to joining. Among the 64 female participants, all but two (who were full-time stay-at-home mothers) had professional careers ranging from 2 to over 15 years in traditional labor markets before moving to platform work. This pattern closely mirrors that of the 44 male participants, all of whom also had established professional careers. Despite these similar career trajectories prior to joining the platform, the data highlights notable disparities in their motivations for participating in platform work.

Surveys indicate that female participants often prioritize family responsibilities when considering participation in platform work, actively seeking opportunities that allow them to balance paid employment with caregiving commitments. In contrast, male participants are primarily motivated by financial objectives, using platform work to supplement or replace their household income. This underscores traditional gender roles, where women frequently serve as primary caregivers, while men experience less influence from family obligations in their decisions regarding platform work. Although financial motivation tends to be less significant for women, it is important to note that those who prioritize caregiving as their main reason for engaging in platform work juggle dual responsibilities: unpaid domestic work alongside additional paid platform work to contribute to household or personal income.

Interestingly, our survey also reveals that flexibility is the least frequently cited primary motivation for entering platform work among both genders. This finding is surprising, as many pundits and scholars view flexibility, whether advertised or genuinely experienced, as a central driver for platform work participation [2, 6, 46, 48].

The following sections discuss how this gendered participation in platform work influences workers' long-term trajectories on digital labor platforms, shaping their perceptions and utilization of platform affordances, as well as how they experience benefits and constraints in relation to their work and careers.

*5.2.2 Flexibility as a Double-Edged Sword for Women: Short-Term Work Empowerment and Long-Term Career Disempowerment.* By synthesizing these findings discussed above with insights from interviews regarding flexibility, we uncovered that the experience of flexibility is gendered, yielding contradictory implications for women's short-term and long-term experiences on digital labor platforms.

Women often view platform-driven flexibility as essential for sustaining their work on digital labor platforms. Interviews show that this appreciation primarily arises from the way platform flexibility allows them to meet their needs related to geographic relocation (e.g., moving to support a partner's career), temporal adjustments (e.g., balancing time between domestic and professional responsibilities), and task management (e.g., engaging in less demanding task while simultaneously taking on caregiving roles).

This finding suggests that flexibility is often interpreted and utilized in ways that reinforce traditional expectations towards women. Therefore, when female workers identify family and caregiving responsibilities as their main motivation for participating in platform work, the underlying narrative highlights their need for flexibility to effectively manage the dual, and sometimes multiple, roles they are required to fulfill as partners, mothers, or daughters.

Ironically, this same flexibility can also constrain their career sustainability, as the need to accommodate multiple roles may limit their availability for better-paying, long-term, and higher-skilled jobs on the platforms. In our study, women reported that they often had to accept smaller, less demanding tasks to manage their domestic responsibilities. This reality positions female workers at the margins of the platform labor market, inherently restricting their opportunities to compete for higher-paying, longer-term, and more skilled work, despite having several years of experience and advanced degrees. These findings provide a nuanced view of the existing literature on the gender pay gap. Notably, previous studies confirm that women typically earn less than men in the platform economy. This income disparity is even more pronounced for women of caregiving age or those with domestic responsibilities [15, 26, 74].

Our longitudinal research advances this discourse by demonstrating that the extended career trajectories of female workers frequently culminate in the economic undervaluation of their work and a sense of career disempowerment. In our survey, female workers consistently reported lower earnings than their male counterparts over the five-year research period. Findings from interviews further reveal that women frequently receive less than half of the financial compensation they used to earn in traditional labor markets. Therefore, they also experience what we term "career disempowerment": a phenomenon whereby prolonged engagement in tasks that fall below the pay range, size, scope, and skill levels, which they could have attained in the traditional workforce or prior to motherhood, undermines their professional confidence and self-identity as career professionals in their fields.

This dual reality highlights a gender-specific challenge in platform-mediated labor: it empowers women with flexibility at the job level; however, when it becomes a career, it may disempower them by undermining their financial security and eroding their professional identity.

*5.2.3 Platform-mediated motherhood penalty: Gendered nature of platformic management.* The concept of motherhood penalty originally refers to the financial and professional disadvantages that mothers face in the traditional workplace, beyond those associated with gender alone. This penalty encompasses both wage reductions and reduced career opportunities that mothers often experience due to societal biases and workplace structures, which assume that caregiving responsibilities detract from women's professional commitment [3, 11, 18, 19]. We introduce the concept of "platform-mediated motherhood penalty", illustrating how digital labor platforms amplify these disadvantages through algorithmic visibility controls, engagement requirements, and rating systems that penalize mothers for caregiving-related breaks.

The CSCW literature has increasingly focused on the distinct challenges that women face, often shaped by platformic management structures [50]. These challenges are documented in our literature [4, 47, 49, 82, 92]. Our research contributes to this discourse by specifically exploring how mothers encounter intensified disadvantages within these platform systems. Drawing on our empirical findings, we discuss how platformic management imposes unique, compounded penalties on mothers, impacting both their immediate work experiences and career sustainability on digital labor platforms.

Platform visibility algorithms systematically disadvantage women, and these biases are magnified for mothers who step away from work to manage caregiving. Women on digital platforms already face reduced visibility in client searches, often appearing lower in rankings than male freelancers [38]. Our findings reveal that for mothers, this visibility bias becomes even more pronounced when caregiving duties require them to take breaks, resulting in a platform-mediated motherhood penalty that diminishes their professional presence and limits job opportunities upon their return. For example, P20, a mother who took time off to care for her children, lost her "top-rated" status, reducing her visibility and making it difficult to re-establish client connections. This reduction in visibility due to algorithmic penalties restricts mothers' access to projects and clients, directly impacting their long-term career viability and contributing to a platform-mediated caregiving penalty.

Our research further demonstrates how platform-imposed requirements for constant connectivity significantly constrain mothers' flexibility, compelling them to manage their professional identities in ways that non-mothers may not experience. Our findings indicate that mothers feel pressured to maintain continuous engagement, fearing that stepping back may harm their professional standing. For instance, P6 described feeling compelled to respond to clients immediately after childbirth, reflecting anxiety that an absence could jeopardize her reputation on the platform. This expectation of constant connectivity enforces a platform-mediated motherhood penalty, where mothers are pressured to prioritize engagement over personal well-being, underscoring the platform's role in intensifying traditional motherhood penalties.

Our findings also reveal that digital platforms intensify the motherhood penalty by evaluating women's life and career paths based on platform-defined metrics of responsiveness and engagement. For mothers, these metrics are especially restrictive, as platform structures penalize caregiving breaks by lowering visibility and engagement scores, thereby eroding their professional standing. For example, P23, who anticipated having multiple children, questioned the sustainability of her career on the platform, as caregiving responsibilities limited her ability to meet the platform's engagement requirements. This marketization of career aspirations forces mothers to compromise between caregiving and professional goals, with extended absences leading to lost competitive standing and reduced financial stability.

By examining visibility algorithms, connectivity requirements, and engagement metrics that intensify traditional motherhood penalties, we highlight the systemic inequities embedded within digital labor platforms. These findings underscore the urgent need for platform designs that recognize and accommodate the diverse realities of women and other users.

## 5.3 Implications for CSCW

Our study contributes to the growing body of CSCW literature by offering gender-inclusive insights into the long-term sustainability of platform work, addressing both women's and men's experiences. We identify implications for platform design and future research that could mitigate gendered challenges and enhance career sustainability on digital labor platforms.

*5.3.1 Redesigning Platform User Experiences for Mitigating Platform-Mediated Motherhood Penalty.* Many challenges that hinder sustainable platform work experiences arise from gendered dynamics. To effectively address these gender-specific issues, future CSCW research and design practices should pay greater attention to the following areas of user experience and interface (re)design.

*Designing for Fair Visibility:* One of the primary challenges women face on digital labor platforms is balancing caregiving responsibilities with the platform's demand for constant availability. Current platform algorithms often penalize freelancers for temporary unavailability, reducing their platform status such as Talent Badges[3] or even rendering their profiles inactive. This has a direct impact on female workers' job opportunities and, in turn, their income. To address this, platforms could implement caregiving-sensitive algorithms that recognize periods of caregiving and accommodate freelancers' limited availability without negatively impacting their rankings.

Also, instead of reducing a freelancer status due to inactivity, the platform could maintain their visibility during caregiving periods by introducing a temporary "active but caregiving" status. This status would signal both to the algorithm and potential clients that the freelancer is available but may require more flexibility in response times or project deadlines. Such adjustments would ensure that caregiving freelancers are not penalized for managing personal responsibilities, allowing them to remain competitive in the marketplace. Additionally, the introduction of "caregiving status" badges offers a transparent solution. Freelancers could activate these badges during caregiving periods, giving clients a clear indication of their availability constraints. This system fosters mutual understanding between clients and freelancers, enabling clients to adjust their expectations regarding response times and deadlines. The result is a more inclusive platform environment that accommodates the real-life demands of freelancers while maintaining professional standards.

[3] https://support.upwork.com/hc/en-us/articles/360049702614-Upwork-s-Talent-Badges

*Designing for Fair Evaluation:* Additionally, platforms should revise their client feedback systems to allow for more nuanced evaluations. A more diverse and flexible evaluation framework would encourage clients to assess the quality of work and adherence to project requirements separately from factors like availability or response times. By focusing on deliverables and project outcomes, this system would offer a more accurate reflection of a freelancer's abilities and contributions. It would also ensure that caregiving-related disruptions do not unfairly affect a freelancer's ratings, promoting a fairer assessment process. Such changes would support the long-term sustainability of platform careers by balancing personal responsibilities with professional excellence, creating a more equitable environment for all workers.

Our data reveals a recurring pattern of female participants predominantly taking on caregiving responsibilities, likely reflecting traditional gender norms. However, it is important to emphasize that caregiving is not inherently gender-specific and can be undertaken by individuals of any gender. Therefore, the insights discussed in this section present design opportunities for the CSCW community to promote more inclusive and equitable work experiences that support individuals of all genders in managing the intersection of caregiving and work.

*5.3.2 Further Research to Improve Career Sustainability.* While our study identifies significant gendered challenges in platform work, further research is needed to deepen our understanding of how to support long-term career sustainability for all genders. We discuss two directions for future research.

*Intersectional Analysis:* Gender constitutes one thread within the broader tapestry of workers' multifaceted identities, shaping how they interact with, are perceived by, and are positioned in platform-mediated work. Building on the CSCW community's longstanding commitment to intersectional approaches, future research should continue to examine how gender intersects with other identity dimensions, such as race, age, and cultural backgrounds, socioeconomic status, to address both the everyday realities and long-term trajectories of platform-mediated work. For example, in traditional labor markets, Glauber (2007) identified multiple intersecting factors that mediate gendered work experiences [34]. Glauber's study shows that gender, race, ethnicity, and marital status collectively shape the motherhood wage penalty, resulting in divergent economic trajectories for African American, Hispanic, and White mothers. In the context of platform labor, Munoz and colleagues (2024) show that digital labor platforms' technological arrangements, such as worker search filters and profile requirements, interact with gender, race, and ethnicity, resulting in disproportionately disadvantaged work experiences for workers from traditionally marginalized backgrounds [70]. Situating this prior evidence alongside our findings, it becomes clear that an intersectional lens can further illuminate how gendered experiences are entangled with other dimensions of identity. Thus, we emphasize the need for future research to adopt longitudinal, intersectional approaches in order to better understand how gendered work experiences are embedded within broader identity dimensions, and how such experiences accrue, persist, or transform over time in platform-mediated work.

*Cross-Platform Analysis:* Our study focuses on a single platform, yet examining different platform types and categories of platform work (e.g., care work, food delivery, beauty services) that may face similar challenges could provide deeper insights into gendered work and career experiences [53,69]. Zheng and colleagues (2024) suggest a declining motherhood penalty in certain segments of China's gig economy, such as among live streamers, ride-share drivers, and food delivery workers, our findings show that platform-mediated motherhood penalties persist, and even deepen, over time in the context of the U.S.-based online freelancing [97]. This divergence may stem from fundamental differences in the nature of platform work. Task-based gig work often offers greater entry-level flexibility, lower skill barriers, and more immediate earnings, which can help mothers maintain labor force attachment. In contrast, knowledge-intensive freelance work typically involves longer project cycles, higher skill thresholds, and sustained client relationships, all of which are more susceptible to disruption from caregiving responsibilities. Whereas Zheng and colleagues highlight flexibility as an equalizing force, we find that flexibility in knowledge work platforms often functions as a short-term enabler but a long-term constraint, especially when platform design mechanisms (e.g., algorithmic visibility and constant availability expectations) penalize caregiving-related interruptions. These contrasting findings underscore the need for cross-platform studies on motherhood penalties in platform work, studies that account for differences in platform type, task structure, and work modality. A comparative analysis of unique or overlapping challenges across these platforms would allow researchers to identify more nuanced gender-specific issues and implications, supporting long-term career sustainability across genders in digital labor platforms.

## 6 CONCLUSIONS

We contribute to the discourse on gender inclusivity within CSCW by enhancing our understanding of gendered work and career trajectories on digital labor platforms and their evolution over time. Our five-year, mixed-method investigation offers three major contributions, advancing both CSCW scholarship and practice. First, empirical insights into gendered platform work and career sustainability: Our survey and interview findings reveal long-term gender disparities in platform work, while our interviews discover how specific platform affordances impact career sustainability for male and female workers. Second, conceptual contributions to understanding gendered platform experiences: We introduce two concepts, career disempowerment and the platform-mediated motherhood penalty, which capture the nuanced ways in which platform structures affect women's work and career experiences. Third, design and research implications for gender-inclusive work environments: Based on these findings, we propose two design and two research implications to guide CSCW research and practice, fostering inclusivity and sustainability in digital labor environments for all genders.

## ACKNOLEDGEMENT

This research is based upon work supported by the National Science Foundation under grant nos. 1665386, 2121624, and 2121638. This research is also supported, in part, by grants from Syracuse University's Office of the Vice President of Research and SOURCE. Any opinions, findings, and conclusions or recommendations expressed in this material are those of the author(s) and do not necessarily reflect the views of the organizations supporting our work.

## REFERENCES

[1] Joan Acker. 1990. Hierarchies, jobs, bodies:: A theory of gendered organizations. *Gend. Soc.* 4, 2 (June 1990), 139–158. https://doi.org/10.1177/089124390004002002

[2] Juan Carlos Alvarez de la Vega, Marta E. Cecchinato, John Rooksby, and Joseph Newbold. 2023. Understanding Platform Mediated Work-Life: A Diary Study with Gig Economy Freelancers. *Proc. ACM Hum.-Comput. Interact.* 7, CSCW1 (April 2023), 1–32. https://doi.org/10.1145/3579539

[3] Deborah J. Anderson, Melissa Binder, and Kate Krause. 2003. The motherhood wage penalty revisited: Experience, heterogeneity, work effort, and work-schedule flexibility. *Ind. Labor Relat. Rev.* 56, 2 (January 2003), 273. https://doi.org/10.2307/3590938

[4] Ira Anjali Anwar, Joyojeet Pal, and Julie Hui. 2021. Watched, but moving: Platformization of beauty work and its gendered mechanisms of control. *Proc. ACM Hum. Comput. Interact.* 4, CSCW3 (January 2021), 1–20. https://doi.org/10.1145/3432949

[5] Michael B. Arthur, Douglas T. Hall, and Barbara S. Lawrence. 1989. Generating new directions in career theory: the case for a transdisciplinary approach. In *Handbook of Career Theory*. Cambridge University Press, 7–25. https://doi.org/10.1017/cbo9780511625459.003

[6] U. Bajwa. 2018. Towards an understanding of workers ' experiences in the global gig economy. *Globalization and Health* 14, (2018), 2–4.

[7] Stephen R. Barley, William H. Dutton, Sara Kiesler, Paul Resnick, Robert E. Kraut, and Joanne Yates. 2004. Does CSCW need organization theory? In *Proceedings of the 2004 ACM conference on Computer supported cooperative work*, November 06, 2004. ACM, New York, NY, USA. https://doi.org/10.1145/1031607.1031628

[8] Arianne Renan Barzilay. 2019. The Technologies of Discrimination: How Platforms Cultivate Gender Inequality. *The Law & Ethics of Human Rights* 13, 2 (November 2019), 179–202. https://doi.org/10.1515/lehr-2019-2006

[9] Allie Blaising, Yasmine Kotturi, Chinmay Kulkarni, and Laura Dabbish. 2021. Making it Work, or Not: A Longitudinal Study of Career Trajectories Among Online Freelancers. *Proc. ACM Hum.-Comput. Interact.* 4, CSCW3 (January 2021), 226:1–226:29. https://doi.org/10.1145/3432925

[10] Virginia Braun and Victoria Clarke. 2012. Thematic analysis. In *APA handbook of research methods in psychology, Vol 2: Research designs: Quantitative, qualitative, neuropsychological, and biological.* American Psychological Association, Washington, 57–71. https://doi.org/10.1037/13620-004

[11] Michelle J. Budig and Paula England. 2001. The wage penalty for motherhood. *Am. Sociol. Rev.* 66, 2 (April 2001), 204. https://doi.org/10.2307/2657415

[12] Judith Butler. 2006. *Gender trouble: Feminism and the subversion of identity.* Routledge, London, England. Retrieved July 22, 2024 from https://www.amazon.com/Gender-Trouble-Feminism-Subversion-Routledge/dp/0415389550

[13] Giulia Campaioli. 2023. Hiding the gender binary behind the "other." *AG About Gender - International Journal of Gender Studies 12.* https://doi.org/10.15167/2279-5057/AG2023.12.23.2112

[14] Tachia Chin, Genyi Li, Hao Jiao, Frederick Addo, and I. M. Jawahar. 2019. Career sustainability during manufacturing innovation: A review, a conceptual framework and future research agenda. *Career Dev. Int.* 24, 6 (October 2019), 509–528. https://doi.org/10.1108/cdi-02-2019-0034

[15] Brendan Churchill. 2023. The gender pay platform gap during the COVID-19 pandemic and the role of platform gender segregation in Australia. *New Technol. Work Employ.* (November 2023). https://doi.org/10.1111/ntwe.12281

[16] Brendan Churchill and Lyn Craig. 2019. Gender in the gig economy: Men and women using digital platforms to secure work in Australia. *J. Sociol.* 55, 4 (December 2019), 741–761. https://doi.org/10.1177/1440783319894060

[17] A. Collin and R. A. Young. 2000. *The future of career.* Cambridge University Press, Cambridge, England. Retrieved from https://www.google.com/books/edition/The_Future_of_Career/NGQA5OFHNbkC?hl=en&gbpv=1&dq=the+concept+of+career+&pg=PR7&printsec=frontcover

[18] Shelley J. Correll, Stephen Benard, and In Paik. 2007. Getting a job: Is there a motherhood penalty? *Am. J. Sociol.* 112, 5 (March 2007), 1297–1339. https://doi.org/10.1086/511799

[19] Ann Crittenden. 2002. *Price of motherhood: Why the Most Important Job in the World is Still the Least Valued.* St Martin's Press, New York, NY. Retrieved from https://www.google.com/books/edition/The_Price_of_Motherhood/Y437K6-FEwcC?hl=en

[20] Nathaniel Ming Curran. 2020. Intersectional English(es) and the Gig Economy: Teaching English Online. *Int. J. Commun. Syst.* 14, 0 (May 2020), 20. Retrieved May 11, 2023 from https://ijoc.org/index.php/ijoc/article/view/11310

[21] Namita Datta, Chen Rong, Sunamika Singh, Clara Stinshoff, Nadina Iacob, Natnael Simachew Nigatu, Mpumelelo Nxumalo, and Luka Klimaviciute. 2023. *Working without borders: The promise and peril of online gig work.* Washington, DC: World Bank. https://doi.org/10.1596/40066

[22] Meredith Dedema and Howard Rosenbaum. 2024. Socio-technical issues in the platform-mediated gig economy: A systematic literature review: An Annual Review of Information Science and Technology (ARIST) paper. *J. Assoc.*

*Inf. Sci. Technol.* 75, 3 (March 2024), 344–374. https://doi.org/10.1002/asi.24868

[23] Sara De Hauw and Jeffrey H. Greenhaus. 2015. Building a sustainable career: the role of work–home balance in career decision making. *Chapters* (2015), 223–238. Retrieved August 6, 2024 from https://ideas.repec.org/h/elg/eechap/15416_15.html

[24] Pelin Demirel, Ekaterina Nemkova, and Rebecca Taylor. 2021. Reproducing Global Inequalities in the Online Labour Market: Valuing Capital in the Design Field. *Work Employ. Soc.* 35, 5 (October 2021), 914–930. https://doi.org/10.1177/0950017020942447

[25] Ans De Vos, Beatrice I. J. M. Van der Heijden, and Jos Akkermans. 2020. Sustainable careers: Towards a conceptual model. *J. Vocat. Behav.* 117, (March 2020), 103196. https://doi.org/10.1016/j.jvb.2018.06.011

[26] Sofia Dokuka, Anastasia Kapuza, Mikhail Sverdlov, and Timofey Yalov. 2022. Women in gig economy work less in the evenings. *Sci. Rep.* 12, 1 (May 2022), 8502. https://doi.org/10.1038/s41598-022-12558-x

[27] William E. Donald, Yehuda Baruch, and Melanie J. Ashleigh. 2019. Striving for sustainable graduate careers: Conceptualization via career ecosystems and the new psychological contract. *Career Dev. Int.* 25, 2 (December 2019), 90–110. https://doi.org/10.1108/cdi-03-2019-0079

[28] Michael Dunn, Isabel Munoz, and Steve Sawyer. 2021. Gender Differences and Lost Flexibility in Online Freelancing During the COVID-19 Pandemic. *Frontiers in Sociology* 6, (August 2021), 738024. https://doi.org/10.3389/fsoc.2021.738024

[29] Johan Espinoza-Rojas, Ignacio Siles, and Thomas Castelain. 2023. How using various platforms shapes awareness of algorithms. *Behav. Inf. Technol.* 42, 9 (July 2023), 1422–1433. https://doi.org/10.1080/0144929x.2022.2078224

[30] Eureka Foong and Elizabeth Gerber. 2021. Understanding Gender Differences in Pricing Strategies in Online Labor Marketplaces. In *Proceedings of the 2021 CHI Conference on Human Factors in Computing Systems* (*CHI '21*), May 07, 2021. Association for Computing Machinery, New York, NY, USA, 1–16. https://doi.org/10.1145/3411764.3445636

[31] Eureka Foong, Nicholas Vincent, Brent Hecht, and Elizabeth M. Gerber. 2018. Women (Still) Ask For Less: Gender Differences in Hourly Rate in an Online Labor Marketplace. *Proc. ACM Hum. Comput. Interact.* 2, CSCW (November 2018), 1–21. https://doi.org/10.1145/3274322

[32] Hernan Galperin. 2021. "This Gig Is Not for Women": Gender Stereotyping in Online Hiring. *Soc. Sci. Comput. Rev.* 39, 6 (December 2021), 1089–1107. https://doi.org/10.1177/0894439319895757

[33] Barney G. Glaser and Anselm L. Strauss. 2000. *Discovery of grounded theory: Strategies for qualitative research.* Routledge, London, England. Retrieved October 28, 2024 from https://www.amazon.com/Discovery-Grounded-Theory-Strategies-Qualitative/dp/1138535168

[34] R. Glauber. 2007. Marriage and the motherhood wage penalty among African Americans, Hispanics, and Whites. *Journal of Marriage and Family* 69, 4 (November 2007), 951–961. https://doi.org/10.1111/J.1741-3737.2007.00423.X

[35] Jeffrey H. Greenhaus and Gerard A. Callanan. 2022. *Advanced introduction to sustainable careers.* Edward Elgar Publishing, Cheltenham, England.

[36] Jeffrey H. Greenhaus, Gerard A. Callanan, and Gary N. Powell. 2024. Advancing research on career sustainability. *J. Career Dev.* 51, 4 (August 2024), 478–497. https://doi.org/10.1177/08948453241260871

[37] Shruti Gupta. 2020. Gendered Gigs: Understanding the gig economy in New Delhi from a gendered perspective. In *Proceedings of the 2020 International Conference on Information and Communication Technologies and Development* (*ICTD '20*), June 17, 2020. Association for Computing Machinery, New York, NY, USA, 1–10. https://doi.org/10.1145/3392561.3394635

[38] Anikó Hannák, Claudia Wagner, David Garcia, Alan Mislove, Markus Strohmaier, and Christo Wilson. 2017. Bias in Online Freelance Marketplaces: Evidence from TaskRabbit and Fiverr. In *Proceedings of the 2017 ACM Conference on Computer Supported Cooperative Work and Social Computing* (*CSCW '17*), February 25, 2017. Association for Computing Machinery, New York, NY, USA, 1914–1933. https://doi.org/10.1145/2998181.2998327

[39] Self-employed, total (% of total employment) (modeled ILO estimate). *World Bank.* Retrieved May 14, 2024 from https://data.worldbank.org/indicator/SL.EMP.SELF.ZS?end=2021&start=2011&view=chart

[40] De Hauw and S. Greenhaus. 2015. Building a sustainable career: The role of work-home balance in career decision making. In *Handbook of research on sustainable careers*, De Vos and A. Van Der Heijden (eds.). Elgar, 223–238.

[41] Clem Herman and Suzan Lewis. 2012. Entitled to a sustainable career? Motherhood in science, engineering, and technology: Entitled to a sustainable career? *J. Soc. Issues* 68, 4 (December 2012), 767–789. https://doi.org/10.1111/j.1540-4560.2012.01775.x

[42] Andrea M. Herrmann, Petra M. Zaal, Maryse M. H. Chappin, Brita Schemmann, and Amelie Lühmann. 2023. "We don't need no (higher) education" - How the gig economy challenges the education-income paradigm. *Technol. Forecast. Soc. Change* 186, (January 2023), 122136. https://doi.org/10.1016/j.techfore.2022.122136

[43] Peter A. Heslin, Lauren A. Keating, and Susan J. Ashford. 2020. How being in learning mode may enable a sustainable career across the lifespan. *J. Vocat. Behav.* 117, 103324 (March 2020), 103324.

https://doi.org/10.1016/j.jvb.2019.103324
[44] Jessica Huang, Ning F. Ma, Veronica A. Rivera, Tabreek Somani, Patrick Yung Kang Lee, Joanna Mcgrenere, and Dongwook Yoon. 2024. Design Tensions in Online Freelancing Platforms: Using Speculative Participatory Design to Support Freelancers' Relationships with Clients. *Proc. ACM Hum.-Comput. Interact.* 8, CSCW1 (April 2024), 1–28. https://doi.org/10.1145/3653700
[45] Keman Huang, Jinhui Yao, and Ming Yin. 2019. Understanding the Skill Provision in Gig Economy from A Network Perspective: A Case Study of Fiverr. *Proc. ACM Hum.-Comput. Interact.* 3, CSCW (November 2019), 132:1–132:23. https://doi.org/10.1145/3359234
[46] Nura Jabagi, Anne-Marie Croteau, Luc K. Audebrand, and Josianne Marsan. 2019. Gig-workers' motivation: thinking beyond carrots and sticks. *Journal of Managerial Psychology; Bradford* 34, 4 (2019), 192–213. https://doi.org/10.1108/JMP-06-2018-0255
[47] Farnaz Jahanbakhsh, Justin Cranshaw, Scott Counts, Walter S. Lasecki, and Kori Inkpen. 2020. An Experimental Study of Bias in Platform Worker Ratings: The Role of Performance Quality and Gender. In *Proceedings of the 2020 CHI Conference on Human Factors in Computing Systems* (*CHI '20*), April 23, 2020. Association for Computing Machinery, New York, NY, USA, 1–13. https://doi.org/10.1145/3313831.3376860
[48] A. James. 2017. *Juggling Work, Home and Family in the Knowledge Economy. Work-Life Advantage.*
[49] Al James. 2022. Women in the gig economy: feminising "digital labour." *Work in the Global Economy* 2, 1 (July 2022), 2–26. https://doi.org/10.1332/273241721x16448410652000
[50] Mohammad Hossein Jarrahi, Will Sutherland, Sarah Beth Nelson, and Steve Sawyer. 2020. Platformic Management, Boundary Resources for Gig Work, and Worker Autonomy. *Comput. Support. Coop. Work* 29, 1-2 (April 2020), 153–189. https://doi.org/10.1007/s10606-019-09368-7
[51] Rosabeth Moss Kanter. 1993. *Men and women of the corporation: New edition.* Basic Books, London, England. Retrieved October 28, 2024 from https://www.amazon.com/Women-Corporation-Rosabeth-Moss-Kanter/dp/0465044549
[52] Pyeonghwa Kim, Charis Asante-Agyei, Isabel Munoz, Michael Dunn, and Steve Sawyer. 2025. Decoding the Meaning of Success on Digital Labor Platforms: Worker-Centered Perspectives. *Proc. ACM Hum.-Comput. Interact.*, 9, 2, Article CSCW059 (April 2025), 29 pages, https://doi.org/10.1145/3710957
[53] Pyeonghwa Kim and Steve Sawyer. 2024. Occupational Diversity in Platform Work: A Comparative Study. In Computer Supported Cooperative Work and Social Computing, November 09, 2024. ACM, New York, NY, USA. https://doi.org/10.1145/3678884.3681853
[54] Pyeonghwa Kim, Eunjeong Cheon, and Steve Sawyer. 2023. Online Freelancing on Digital Labor Platforms: A Scoping Review. In *Companion Publication of the 2023 Conference on Computer Supported Cooperative Work and Social Computing* (*CSCW '23 Companion*), October 14, 2023. Association for Computing Machinery, New York, NY, USA, 259–266. https://doi.org/10.1145/3584931.3607011
[55] Pyeonghwa Kim and Steve Sawyer. 2023. Many Futures of Work and Skill: Heterogeneity in Skill Building Experiences on Digital Labor Platforms. In *Proceedings of the 2nd Annual Meeting of the Symposium on Human-Computer Interaction for Work* (*CHIWORK '23*), September 20, 2023. Association for Computing Machinery, New York, NY, USA, 1–9. https://doi.org/10.1145/3596671.3597655
[56] E. E. Kossek, M. Valcour, and P. Lirio. 2014. Organizational strategies for promoting work-life balance and wellbeing. *Work and wellbeing* 3, (2014), 295–319.
[57] Ellen Ernst Kossek, Monique Valcour, and Pamela Lirio. 2014. The sustainable workforce: Organizational strategies for promoting work-life balance and wellbeing. *Wellbeing*, 1–24. https://doi.org/10.1002/9781118539415.wbwell030
[58] Airi Lampinen, Christoph Lutz, Gemma Newlands, Ann Light, and Nicole Immorlica. 2018. Power Struggles in the Digital Economy: Platforms, Workers, and Markets. In *Companion of the 2018 ACM Conference on Computer Supported Cooperative Work and Social Computing* (*CSCW '18 Companion*), October 30, 2018. Association for Computing Machinery, New York, NY, USA, 417–423. https://doi.org/10.1145/3272973.3273004
[59] Barbara S. Lawrence, Douglas T. Hall, and Michael B. Arthur. 2015. Sustainable careers then and now. In *Handbook of Research on Sustainable Careers.* Edward Elgar Publishing. https://doi.org/10.4337/9781782547037.00033
[60] Min Kyung Lee, Daniel Kusbit, Evan Metsky, and Laura Dabbish. 2015. Working with Machines: The Impact of Algorithmic and Data-Driven Management on Human Workers. In *Proceedings of the 33rd Annual ACM Conference on Human Factors in Computing Systems* (*CHI '15*), April 18, 2015. Association for Computing Machinery, New York, NY, USA, 1603–1612. https://doi.org/10.1145/2702123.2702548
[61] Ming D. Leung and Sharon Koppman. 2018. Taking a Pass: How Proportional Prejudice and Decisions Not to Hire Reproduce Gender Segregation. *American Journal of Sociology 124*, 762–813. https://doi.org/10.1086/700677
[62] Anna Lindqvist, Marie Gustafsson Sendén, and Emma A. Renström. 2021. What is gender, anyway: a review of the options for operationalising gender. *Psychol. Sex.* 12, 4 (October 2021), 332–344. https://doi.org/10.1080/19419899.2020.1729844
[63] Paul Luff, Jon Hindmarsh, and Christian Heath. 2000. *Workplace Studies: Recovering Work Practice and Informing*

*System Design*. Cambridge University Press. Retrieved from https://play.google.com/store/books/details?id=w5S8sLDhAqQC
[64] Ning F. Ma, Veronica A. Rivera, Zheng Yao, and Dongwook Yoon. 2022. "Brush it Off": How Women Workers Manage and Cope with Bias and Harassment in Gender-agnostic Gig Platforms. In *Proceedings of the 2022 CHI Conference on Human Factors in Computing Systems* (*CHI '22*), April 29, 2022. Association for Computing Machinery, New York, NY, USA, 1–13. https://doi.org/10.1145/3491102.3517524
[65] Shuhao Ma, Valentina Nisi, John Zimmerman, and Nuno Nunes. 2023. Mapping the Research Landscape of the Gig Work for Design on Labour Research. In *IASDR Conference Series*, 2023. . https://doi.org/10.21606/iasdr.2023.473
[66] Jeroen Meijerink and Anne Keegan. 2025. Careers and the gig economy: analyzing the broader effects of gig work on career patterns, dynamics and outcomes. In *Research Handbook of Careers in the Gig Economy*. Edward Elgar Publishing, 151–166. https://doi.org/10.4337/9781035318537.00017
[67] Jarrahi Mohammad Hossein and Will Sutherland. 2019. Algorithmic management and algorithmic competencies: Understanding and appropriating algorithms in gig work. In *International Conference on Information*. Springer, 578–589.
[68] Mareike Mohlmann and Lior Zalmanson. 2017. Hands on the Wheel: Navigating Algorithmic Management and Uber Drivers' Autonomy. In *ICIS 2017 Proceedings*, 2017. . Retrieved April 3, 2024 from https://aisel.aisnet.org/icis2017/DigitalPlatforms/Presentations/3/
[69] Isabel Munoz, Michael Dunn, and Steve Sawyer. 2022. Flexibility, Occupation and Gender: Insights from a Panel Study of Online Freelancers. In *Information for a Better World: Shaping the Global Future* (*Lecture Notes in Computer Science*), 2022. Springer International Publishing, Cham, 311–318. https://doi.org/10.1007/978-3-030-96957-8_27
[70] Isabel Munoz, Pyeonghwa Kim, Clea O'Neil, Michael Dunn, and Steve Sawyer. 2024. Platformization of inequality: Gender and race in digital labor platforms. *Proc. ACM Hum. Comput. Interact.* 8, CSCW1 (April 2024), 1–22. https://doi.org/10.1145/3637385
[71] Isabel Munoz, Steve Sawyer, and Michael Dunn. 2022. New futures of work or continued marginalization? The rise of online freelance work and digital platforms. In *Proceedings of the 1st Annual Meeting of the Symposium on Human-Computer Interaction for Work* (*CHIWORK 2022*), June 08, 2022. Association for Computing Machinery, New York, NY, USA, 1–7. https://doi.org/10.1145/3533406.3533412
[72] Karen L. Newman. 2011. Sustainable careers. *Organ. Dyn.* 40, 2 (April 2011), 136–143. https://doi.org/10.1016/j.orgdyn.2011.01.008
[73] Payoneer. 2023. *2023 Freelancer Insights Report*. Retrieved from https://pubs.payoneer.com/docs/2023-freelancer-insights-report.pdf
[74] Robert A. Peterson. 2022. Heterogeneity in the US gig economy with a focus on gender. *International Journal of Applied Decision Sciences* 15, 3 (January 2022), 365–384. https://doi.org/10.1504/IJADS.2022.122641
[75] M. Anand Shankar Raja, A. V. Akshay Kumar, Neha Makkar, Senthil Kumar, and S. Bhargav Varma. 2022. The Future of the Gig Professionals: A Study Considering Gen Y, Gen C, and Gen Alpha. In *Sustainability in the Gig Economy: Perspectives, Challenges and Opportunities in Industry 4.0*, Ashish Gupta, Tavishi Tewary and Badri Narayanan Gopalakrishnan (eds.). Springer Nature Singapore, Singapore, 305–324. https://doi.org/10.1007/978-981-16-8406-7_23
[76] Noopur Raval and Joyojeet Pal. 2019. Making a "Pro": "Professionalism" after Platforms in Beauty-work. *Proc. ACM Hum.-Comput. Interact.* 3, CSCW (November 2019), 1–17. https://doi.org/10.1145/3359277
[77] Alex Rosenblat. 2018. *Uberland: How Algorithms Are Rewriting the Rules of Work*. Univ of California Press. https://doi.org/10.1525/9780520970632
[78] Linda Schweitzer, Sean Lyons, and Chelsie J. Smith. 2023. Career Sustainability: Framing the Past to Adapt in the Present for a Sustainable Future. *Sustain. Sci. Pract. Policy* 15, 15 (July 2023), 11800. https://doi.org/10.3390/su151511800
[79] Jason Seawright. 2016. *Strategies for social inquiry: Multi-method social science: Combining qualitative and quantitative tools: Combining qualitative and quantitative tools*. Cambridge University Press, Cambridge, England. https://doi.org/10.1017/cbo9781316160831
[80] Andrey Shevchuk and Denis Strebkov. 2018. Safeguards against opportunism in freelance contracting on the internet. *Br. J. Ind. Relat.* 56, 2 (June 2018), 342–369. https://doi.org/10.1111/bjir.12283
[81] A. Strauss and J. Corbin. 1998. *Basics of qualitative research: Techniques and procedures for developing grounded theory*. Sage Publications, Thousand Oaks, California.
[82] Elisabetta Stringhi. 2022. Addressing gendered affordances of the platform economy: The case of UpWork. *Internet Policy Review: Journal on Internet Regulation* 11, 1 (March 2022), 1–28. https://doi.org/10.14763/2022.1.1634
[83] Simone Stumpf, Anicia Peters, Shaowen Bardzell, Margaret Burnett, Daniela Busse, Jessica Cauchard, and Elizabeth Churchill. 2020. Gender-inclusive HCI research and design: A conceptual review. *Found. Trends® Hum.–Comput. Interact.* 13, 1 (2020), 1–69. https://doi.org/10.1561/1100000056
[84] Konstantinos Stylianou. 2016. *Terms of Service and Human Rights: an Analysis of Online Platform Contracts*. Editora

Revan. Retrieved April 3, 2024 from https://eprints.gla.ac.uk/299823/

[85] Sherry E. Sullivan and Yehuda Baruch. 2009. Advances in Career Theory and Research: A Critical Review and Agenda for Future Exploration. *J. Manage.* 35, 6 (December 2009), 1542–1571. https://doi.org/10.1177/0149206309350082

[86] Shagini Udayar, Leandro Ivan Canzio, Ieva Urbanaviciute, Jonas Masdonati, and Jérôme Rossier. 2021. Significant life events and career sustainability: A three-wave study. *Sustainability* 13, 23 (November 2021), 13129. https://doi.org/10.3390/su132313129

[87] Monique Valcour. 2015. Facilitating the crafting of sustainable careers in organizations. In *Handbook of Research on Sustainable Careers*. Edward Elgar Publishing. https://doi.org/10.4337/9781782547037.00007

[88] Beatrice I. J. M. Van der Heijden and Ans De Vos. 2015. Sustainable careers: introductory chapter. *Handbook of Research on Sustainable Careers*, 1–19. https://doi.org/10.4337/9781782547037.00006

[89] Rama Adithya Varanasi, Divya Siddarth, Vivek Seshadri, Kalika Bali, and Aditya Vashistha. 2022. Feeling Proud, Feeling Embarrassed: Experiences of Low-income Women with Crowd Work. In *Proceedings of the 2022 CHI Conference on Human Factors in Computing Systems* (*CHI '22*), April 29, 2022. Association for Computing Machinery, New York, NY, USA, 1–18. https://doi.org/10.1145/3491102.3501834

[90] Valentina Vukmirović, Željko Spasenić, and Miloš Milosavljević. 2023. A bibliometric analysis and future research agenda for online labour platforms. *Stanovništvo* 61, 2 (2023), 183–207. Retrieved from http://ebooks.ien.bg.ac.rs/1989/1/STNV_61%282%29_537_Vukmirovi%C4%87-et-al_183-207.pdf

[91] Judy Wajcman. 2000. Reflections on gender and technology studies: *Soc. Stud. Sci.* 30, 3 (June 2000), 447–464. https://doi.org/10.1177/030631200030003005

[92] Natasha A. Webster and Qian Zhang. 2020. Careers Delivered from the Kitchen? Immigrant Women Small-scale Entrepreneurs Working in the Growing Nordic Platform Economy. *NORA - Nordic Journal of Feminist and Gender Research* 28, 2 (April 2020), 113–125. https://doi.org/10.1080/08038740.2020.1714725

[93] Christine L. Williams, Chandra Muller, and Kristine Kilanski. 2012. Gendered organizations in the new economy. *Gend. Soc.* 26, 4 (August 2012), 549–573. https://doi.org/10.1177/0891243212445466

[94] M. Williams and T. Moser. 2019. The art of coding and thematic exploration in qualitative research. *International management review* 15, 1 (2019), 45–55. Retrieved from https://www.academia.edu/download/82465402/imr-v15n1art4.pdf

[95] Alex J. Wood, Vili Lehdonvirta, and Mark Graham. 2018. Workers of the Internet unite? Online freelancer organisation among remote gig economy workers in six Asian and African countries. *New Technology, Work and Employment* 33, 2 (July 2018), 95–112. https://doi.org/10.1111/ntwe.12112

[96] Alex Wood and Vili Lehdonvirta. 2022. *Platforms Disrupting Reputation: Precarity and Recognition Struggles In The Remote Gig Economy*. SSRN. Retrieved from https://play.google.com/store/books/details?id=gZXhzwEACAAJ

[97] Qi Zheng, Zitong Qiu, and Weiguo Yang. 2024. The shifting motherhood penalty and fatherhood premium in China's gig economy: Impact of parental status on income changes. *Int. Labour Rev.* 163, 2 (June 2024), 173–197. https://doi.org/10.1111/ilr.12407